# Broadband high reflectivity in subwavelength-grating slab waveguides

**Xuan Cui,[1] Hao Tian,[1] and Zhongxiang Zhou,[1, *]**

[1] *Department of Physics, Harbin Institute of Technology, Harbin 150001, China*

**Abstract:** We computationally study a subwavelength dielectric grating structure, show that slab waveguide modes can be used to obtain broadband high reflectivity, and analyze how slab waveguide modes influence reflection. A structure showing interference between Fabry-Perot modes, slab waveguide modes, and waveguide array modes is designed with ultra-broadband high reflectivity. Owing to the coupling of guided modes, the region with reflectivity R > 0.99 has an ultra-high bandwidth ($\Delta f / \bar{f} > 30\%$). The incident-angle region with R > 0.99 extends over a range greater than 40°. Moreover, an asymmetric waveguide structure is studied using a semiconductor substrate.

## 1. Introduction

Subwavelength structures are of interest in a wide range of theoretical fields and applications, including cavity quantum electrodynamics, polariton lasers, filters, splitters, and couplers. Recently, high-contrast subwavelength gratings (HCGs) have been used as reflectors in quantum cavities [1], vertical-cavity surface-emitting lasers (VCSELs) [2], and optomechanical nanoresonators [3], replacing the conventional distributed Bragg reflectors because of its remarkable performance in terms of dispersion, reflectivity, and bandwidth [4]. HCG reflectors can introduce ultra-high dispersion to quantum cavities, which can be engineered to control the cavity performance by, for example, modifying phase and group velocities [5], controlling the density of states (DOS) of the photonic modes [6], and creating synthetic magnetic fields [7]. In addition, for tunable etalon-type device applications, such as lasers, filters, and detectors, the tuning range is significantly extended using HCG broadband reflectors. However, HCG reflectors need a sacrifice layer [8] (generally an air gap), which complicates the fabrication of an integrated device, and HCG can hardly be made into a membrane owing to its discontinuous structure.

Recently, a numerical algorithm was developed to simulate rectangular gratings based on the choice of propagation mode-set [9], which converges much faster than the Rigorous Coupled Wave Analysis (RCWA) does for HCG. The results demonstrated that the interference of the first two propagation modes in the grating, which are called waveguide-array (WGA) modes [1, 9], causes the extraordinary features of HCG. The WGA modes are coupled between grating bars and gaps and have a similar dispersion relation with slab waveguide modes. Thus, guided modes might have a similar effect.

In this Letter, we develop a subwavelength-grating slab waveguide structure, which is easier to fabricate for membranes than HCG is. Furthermore, we analyze the mechanism of its high reflectivity, in which the coupling of guided modes plays a significant role. The calculation results show that the interference between the guided modes and Fabry-Perot modes is essential for the transmission properties and broadband reflectivity. An ultra-broadband reflectivity can be obtained via coupling between guided modes, WGA modes, and Fabry-Perot modes. Moreover, this high reflectivity is observed in a large incident-angle region.

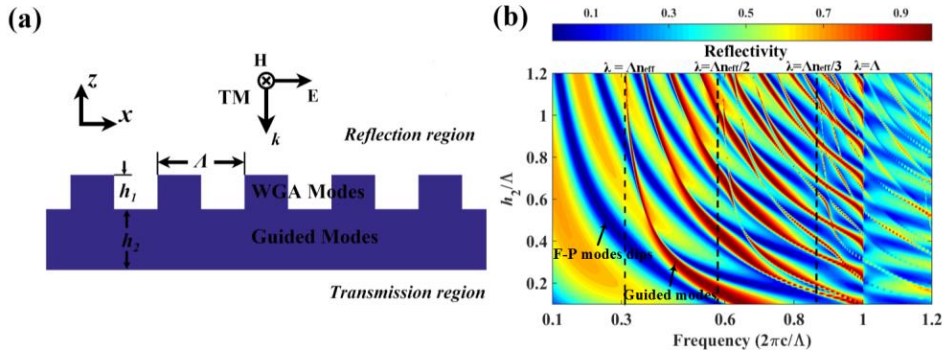

Fig. 1. (a) Schematic of the free-standing grating slab waveguide structure. (b) Reflectivity contour of the structure as a function of frequency and slab-layer thickness simulated using RCWA, with TM-polarized, surface-normal incident waves and the parameters $h_1 = 0.167\Lambda$, $\varepsilon_r = 11.9$, and $\eta = 0.5$. The contour map shows a dispersion shape of guided modes but with some cut-off frequencies (dashed lines), which has affinities with the dispersion of WGA modes. Intricate transmission properties occur in the $\Lambda < \lambda < \Lambda n_{eff}$ range owing to interference.

## 2. Effect of slab waveguide modes

A schematic of our proposed structure is shown in Fig. 1(a). All length units are normalized by the grating period ($\Lambda$) with the following parameters: grating-layer thickness $h_1$, and slab-waveguide-layer thickness $h_2$. Here, guided-mode dispersion introduced by the slab-waveguide layer plays a significant role in transmission. Fig. 1(b) shows the reflectivity contour map against the normalized frequency (by $2\pi c/\Lambda$) and slab-layer thickness for a surface-normal TM-plane wave incident onto the free-standing structure, with the permittivity of both grating and slab material permittivity $\varepsilon_r = 11.9$ (Si), grating thickness $h_1/\Lambda = 0.167$, and a duty cycle of $\eta = 0.5$ (in this paper, $\eta = 0.5$ is always true unless otherwise mentioned) using RCWA [10]. Three regions can be clearly identified with different wave vectors in the propagation direction $k_z$, which is defined as

$$k_z^2 = (2\pi n_i / \lambda)^2 - (2\pi m / \Lambda)^2. \tag{1}$$

where $\lambda$ is the wavelength, $m$ is the diffraction order, and $n_i$ is the refraction index of the material in each layer, which is unity in air.

For $\lambda > \Lambda n_{eff}$ (the left region in Fig. 1(b), where $n_{eff}$ is the effective refractive index of the structure), $k_z$ is always imaginary when $m > 0$. The reflection shows classical Fabry-Perot dispersion. The curve for reflectivity versus frequency and slab thickness displays an approximate $1/f$ line shape. For $\lambda < \Lambda$ (the right region in Fig. 1(b)), $k_z$ can be real in air when $m > 0$. Thus, high-order diffraction exists in the transmission, which reduces the total contrast of the peaks and dips. For $\Lambda < \lambda < \Lambda n_{eff}$, $k_z$ is imaginary in the transmission region (air below), but it can be real in the slab layer when $m > 0$. Therefore, a real $k_x$ (or $k_\parallel$) is introduced into the slab layer. The imaginary $k_z$ in the transmission region ensures that the wave power is entirely confined in the zero-order diffraction. The real $k_z$ and $k_x$ in the slab layer lead to the coupling of guided modes, resulting in intricate transmission properties. The real guided modes introduced by the slab layer have a similar dispersion with WGA modes. The dispersion curves shown in Fig. 1(b) illustrate a clear line shape of guided modes but with different cut-off frequencies for each curve group, compared to WGA modes. These three cut-off lines in the region represent $m$ = 1, 2, 3 ($k_\parallel = 2\pi m / \Lambda$). In each $m$-order diffraction group, different order waveguide modes exist and have the same cut-off lines, similar to the dispersion of ordinary guided modes. The detailed reflection characteristics resemble those of WGA dispersion. Characteristics common with the WGA modes include (1) the cut-off frequency mentioned above; (2) the existence of a so-called "dual mode" regime[11], in which high reflectivity occurs; and (3) the occurrence of crossings (when $m$ = 1, 3 or 0, 2 intersect) and anti-crossings (when $m$ = 0, 1; 1, 2; or 2, 3 intersect) when different $m$-order mode curves intersect. Anti-crossings are usually known to be indicators of strong coupling, which results in broadband high reflectivity or high-Q resonances.

In the present study, the grating-layer thickness is $0.167\Lambda$. Thus, the grating layer can be simply considered to provide the parallel wave vector in the structure via diffraction, which is demonstrated by the dispersion curves of classical guided modes. In this case, the so-called "dual mode" interference occurs between Fabry-Perot modes ($m$ = 0) and the guided modes of high order diffraction ($m > 0$). The dispersion relation shown in Fig. 1(b) illustrates that when the curve of the Fabry-Perot modes (the dark blue region in the contour map) encounters the first-order diffraction mode, anti-crossings occur [11]. Moreover, when these two curves show similar slopes, broadband high reflectivity occurs at the intersection; otherwise, high-Q resonances occur. Fabry-Perot modes are weak-coupling resonances, and when the waveguide-mode curves cross the curves of the Fabry-Perot modes with different slopes, they play the role of discrete resonance, as in Fano resonances [12]. On the other hand, when two types of modes have similar dispersion slopes, the reflected wave phases resulting from the two coupling modes interfere with each other, attenuating the transmission power in a large frequency range. This phenomenon indicates that different dispersion relations can be engineered for various applications by modifying the parameters of the structure.

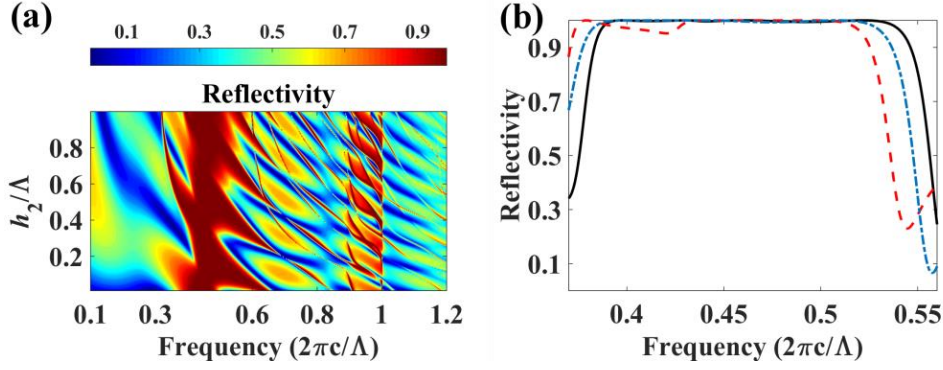

Fig. 2. (a) Reflectivity contour as a function of frequency and grating-layer thickness with $h_1 = 0.685\Lambda$. The thick grating layer can sustain WGA modes, which are essential for HCGs. The interference among guided modes, WGA modes, and Fabry-Perot modes leads to the broadband high reflection (dark red area). (b) Reflectivity with the parameters described in (a) for three different configurations: $h_2 = 0.45\Lambda$ and $\eta = 0.5$ (black line), $h_2 = 0.5\Lambda$ and $\eta = 0.5$ (red dashed line), and $h_2 = 0.45\Lambda$ and $\eta = 0.55$ (blue dash-dotted line).

## 3. Interference between WGA modes and slab waveguide modes

In the case above, the grating layer mainly has two effects: (1) introducing the parallel wave vector through diffraction and (2) modifying the effective index of the entire structure. Because of the small thickness we set, the wave phase produced by the grating layer is not considerable. With increasing grating-layer thickness, the wave phase introduced by the grating cannot be neglected, which will introduce mode coupling in the grating bars (WGA modes). Moreover, the grating layer influences the effective index, which determines the Fabry-Perot dispersion curves. The interference between WGA modes and Fabry-Perot modes can yield broadband high reflectivity, and this interference has been utilized as mirrors in quantum wells [13] and VCSELs. The HCG gratings have the structure shown in Fig. 1(a) without the slab layer. The additional slab layer on the HCG, which introduces the coupling of guided modes, not only modifies the dispersion relation but also improves the reflector performance. The simulation result in Fig. 2(a) shows the reflectivity contour map versus the normalized frequency ($2\pi c/\Lambda$) and slab-layer thickness $h_2$ for a surface-normal incident TM-plane wave, with the material permittivity $\varepsilon_r = 11.9$ (Si) and grating-layer thickness $h_1 = 0.685\Lambda$. Further, two cut-off lines ($\lambda = \Lambda n_{eff}$ and $\lambda = \Lambda$) are observed, which divide the contour map into three parts. As mentioned above, an increasing grating thickness modifies the dispersion curves. Thus, the position and slope of the curves of the Fabry-Perot modes (blue region in the figure) have significantly changed. Moreover, as the grating thickness increases, the WGA modes, which exist in the high-refractive-index bars and air gaps, provide more complicated coupling. As shown in the contour map, when the slab layer is thin (in the bottom area), a high-reflectivity region due to the WGA exists. As the slab-layer thickness increases, guided modes emerge, couple with WGA modes, and significantly expand the high-reflectivity range. In this case, the reflection is determined by interference among the Fabry-Perot modes ($m = 0$), waveguide modes ($m > 0$), and WGA modes (grating contribution). We cannot distinguish guided modes from WGA modes in Fig. 2(a) because of their similar dispersion relations. These three modes affect the wave phase together and lead to ultra-broadband high reflectivity.

Owing to the guided modes, the high-reflectivity range is larger than those attained with only WGA modes as traditional HCG gratings. The black curve in Fig. 2(b) shows the reflectivity at a fixed slab-layer thickness ($h_2 = 0.45\Lambda$). The high-reflectivity (R > 0.99) bandwidth ($\Delta f / \bar{f}$) is greater than 30%. In addition, all the simulations are performed with normalized units. Therefore, the structure can be easily designed by adjusting $\Lambda$ for a certain frequency. For example, for the traditional optical communication range, we set the central wavelength as 1.55 μm; then, the high-reflection range is from 1.31 μm to 1.79 μm, with the structure parameters $\Lambda = 0.698 \mu m$, $h_1 = 0.478$ μm, $h_2 = 0.314$ μm, and $\varepsilon_r = 11.9$. We consider variations of up to 10% in the slab thickness $h_2$ and grating duty circle $\eta$. The red dashed line and blue dash-dotted line show the influence of $h_2$ and $\eta$, respectively. Owing to the complicated interference, there still remains a large range of high reflectivity. In fact, the fabrication technology could control the error within 5% in the parameters for a central wavelength of 1.55 μm. The permittivity here is chosen as that of silicon. Moreover, silicon and many other semiconductor materials have a high refractive index (2.8~3.5) and show little dispersion from the infrared region to the terahertz region. Thus, by modifying $\Lambda$ (about 100 times that at 1.55 μm), this structure can also be used in the terahertz region.

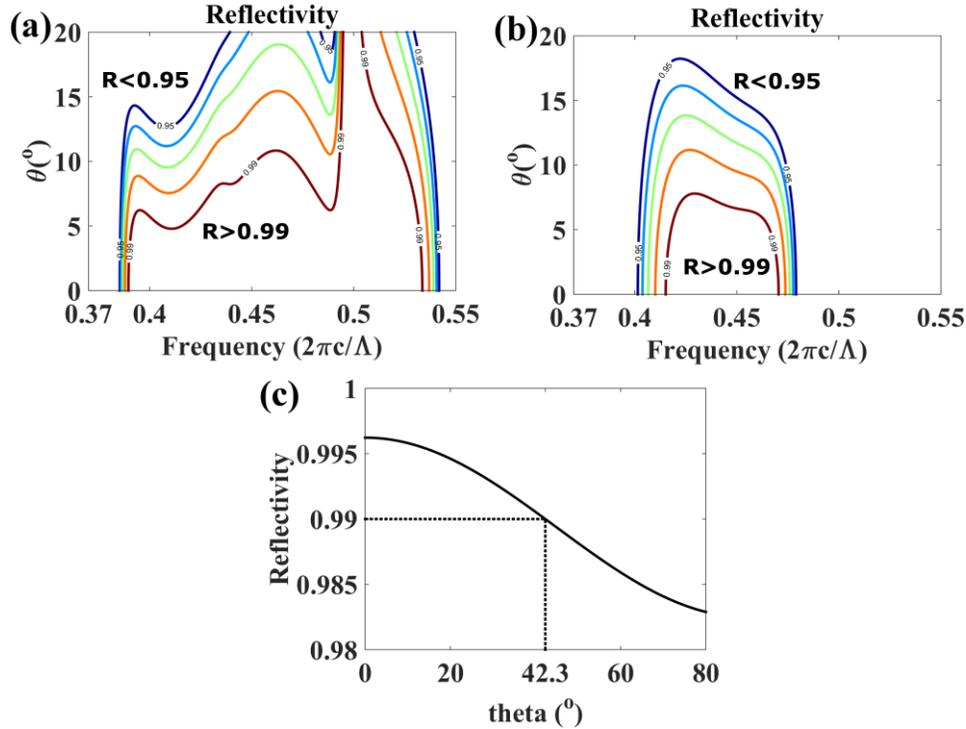

Fig. 3. (a) Reflectivity contour as a function of frequency and incident angle with the parameters $h_1 = 0.685\Lambda$ and $h_2 = 0.45\Lambda$. (b) Reflectivity contour as a function of frequency and incident angle with the parameters $h_1 = 0.685\Lambda$ and $h_2 = 0$ (HCGs). (c) Reflectivity as a function of incident angle at a frequency of $0.5 \times 2\pi c / \Lambda$. The high reflection angle region is significantly expanded because of the slab waveguide modes.

The waveguide modes introduce rather intricate phase interference to the reflection, some of which significantly benefits the structure as a reflector. Fig. 3(a) and Fig. 3(b) show contour maps of the reflectivity as a function of frequency and incident angle at the high-reflectivity frequency region with $h_1 = 0.685\Lambda$, $h_2 = 0.45\Lambda$ and $h_1 = 0.685\Lambda$, $h_2 = 0$, respectively. The contour maps indicate that the angle tolerance greatly increases with the aid of the slab layer, compared to the case of only a WGA (HCGs), owing to the coupling of guided modes. At a frequency of approximately $0.5 \times 2\pi c / \Lambda$ with the structure parameters in Fig. 3(a), the angle tolerance is up to 40°, as shown in Fig. 3(c). This huge incident-angle redundancy provides flexibility in integrating components. In addition, the large phase shift produced by the large angle introduces more dispersion properties of cavity quantum electrodynamics systems using this reflector.

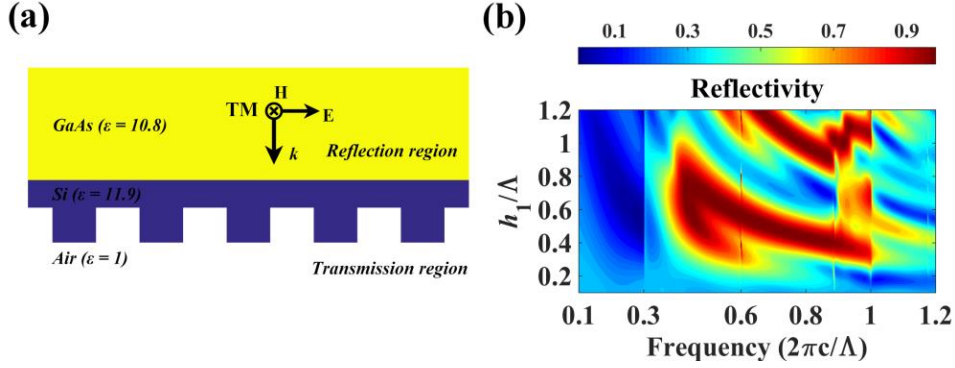

Fig. 4. (a) Schematic of the asymmetric waveguide structure. (b) Reflectivity contour of the structure with a substrate ($\varepsilon_r = 10.9$) as a function of frequency and grating thickness with $h_2 = 0.45\Lambda$.

The transmission properties of the free-standing structure benefit from the symmetric slab waveguide modes. In addition, the asymmetric waveguide modes influence the reflection properties. Fig. 4 shows the reflectivity contour with a substrate of permittivity $\varepsilon_r = 10.8$ (GaAs) as a function of frequency for a grating with parameters $h_1 = 0.685\Lambda$ and $h_2 = 0.45\Lambda$. The normally incident electromagnetic wave propagates from the substrate material to the structure and into air. As shown in Fig. 4, there also exist several high-reflectivity areas and the same cut-off lines in the contour map, which implies that the asymmetric waveguide modes can be exploited to design a reflector. Moreover, this structure with a slab-waveguide layer and substrate layer can be easily used for integrated components through electron-beam deposition, molecular-beam epitaxy, or some other modern membrane-growth technology.

## 4. Conclusion

In summary, we proposed a subwavelength-grating slab waveguide structure, calculated its reflection properties, and analyzed the effect of slab waveguide modes. The interference between guided modes and Fabry-Perot modes lead to complicated reflection properties including broadband high reflectivity. An ultra-broadband ($\Delta f / \bar{f} > 30\%$ with R > 0.99) high reflectivity was observed with phase interference among the guided modes, Fabry-Perot modes, and WGA modes. On introducing the guided modes, the angle tolerance significantly increased, which implies that R > 0.99 is maintained at incident angles greater than 40°. Moreover, we simulated an asymmetric waveguide setup and observed high-reflectivity areas. Because of the guided modes introduced by the slab-waveguide layer, a feasible dispersion can be realized; thus, our structure is promising for use in quantum cavities and integrated components.